\documentstyle[12pt]{article}
\begin{document}
\renewcommand{\thefootnote}{\fnsymbol{footnote}}
\newpage
\pagestyle{empty}
\setcounter{page}{0}



\newcommand{\norm}[1]{{\protect\normalsize{#1}}}
\newcommand{\LAP}
{{\small E}\norm{N}{\large S}{\Large L}{\large A}\norm{P}{\small P}}
\newcommand{\sLAP}{{\scriptsize E}{\footnotesize{N}}{\small S}{\norm L}$
${\small A}{\footnotesize{P}}{\scriptsize P}}
\begin{minipage}{5.2cm}
\begin{center}
{\bf G{\sc\bf roupe d'} A{\sc\bf nnecy}\\
\ \\
Laboratoire d'Annecy-le-Vieux de Physique des Particules}
\end{center}
\end{minipage}
\hfill
\hfill
\begin{minipage}{4.2cm}
\begin{center}
{\bf G{\sc\bf roupe de} L{\sc\bf yon}\\
\ \\
Ecole Normale Sup\'erieure de Lyon}
\end{center}
\end{minipage}
\centerline{\rule{12cm}{.42mm}}

\vspace{20mm}

\begin{center}

{\LARGE {\bf An action for N=4 supersymmetric self-dual Yang-Mills
theory}}\\[1cm]

\vspace{10mm}

{\large E.  Sokatchev$^{1a}$}

{\em Laboratoire de Physique Th\'eorique }\LAP\footnote{URA 14-36
du CNRS, associ\'ee \`a l'Ecole Normale Sup\'erieure de Lyon  et \`a
l'Universit\'e de Savoie

\noindent
$^1$ Groupe d'Annecy: LAPP, Chemin de Bellevue BP 110, F-74941
Annecy-le-Vieux Cedex, France. \\
\noindent
$^a$ e-mail address: sokatche@lapphp0.in2p3.fr

\noindent
}\\

\end{center}
\vspace{20mm}

\centerline{ {\bf Abstract}}
\vskip5mm
\indent
The $N=4$ supersymmetric self-dual Yang-Mills theory in a four-dimensional
space with signature $(2,2)$ is formulated in harmonic superspace. The
on-shell constraints of the theory are reformulated in the equivalent form of
vanishing
curvature conditions for three gauge connections (one harmonic and two
space-time). The constraints are then obtained as variational equations from
a superspace action of the Chern-Simons type.  The action is manifestly
$SO(2,2)$ invariant. It can be viewed as the Lorentz-covariant form of the
light-cone superfield action proposed by Siegel.

\vfill
\rightline{\LAP-A-545/95}
\rightline{ September 1995}

\newpage
\pagestyle{plain}
\renewcommand{\thefootnote}{\arabic{footnote}}

\newpage\setcounter{page}1

\section{Introduction}
The interest in self-dual theories in a
four-dimensional space-time with signature $(2,2)$ has risen considerably
after the observation of Ooguri and Vafa \cite{Vafa} that the string with
local $N=2$ world-sheet supersymmetry has only one state describing
self-dual Yang-Mills (open string) and self-dual gravity (closed string).
Soon afterwards Parkes \cite{Parkes} proposed a field-theory action
\footnote{A similar action for self-dual Yang-Mills had earlier appeared in
a different context in \cite{Leznov}.}, allegedly corresponding to the
amplitudes of this string. This action uses a special Lorentz-non-covariant
gauge for Yang-Mills theory of the type first considered by Yang \cite{Yang},
in which only one degree of freedom is left (as opposed to the three degrees
of freedom in the covariant self-duality condition). Besides the lack of
covariance, this action contradicts standard dimensional counting, as noted in
\cite{Siegel2,Siegel3} (it requires a dimensionful coupling constant which
is not natural in a four-dimensional Yang-Mills theory).

In \cite{Siegel0} Siegel put forward the idea that the $N=4$ string, when
properly
quantized, is in fact equivalent to the $N=2$ one. He further argued that the
corresponding field theory is actually $N=4$ supersymmetric self-dual
Yang-Mills
(SSDYM) in the open case or $N=8$ self-dual supergravity in the closed case.
In \cite{Siegel3} he also presented a Green-Schwarz-type formulation of that
string. \footnote{Some comments on the
validity of the latter have recently been made in \cite{Pope}.}

In general, the multiplet of
$N$-extended SSDYM theory contains helicities from $+1$ to $1-{N\over 2}$. So,
it includes all the helicities
from $+1$ to $-1$ in the maximal case $N=4$ only. As a consequence, in the
latter case the degrees of freedom appear in Lagrangian pairs and one is able
to write down a Lorentz-covariant action for the theory. So far this action
has been presented either in component (i.e., not manifestly
supersymmetric) \cite{Siegel2,Gates} or light-cone superspace (i.e., not
manifestly Lorentz-covariant) form \cite{Siegel1,Siegel2}. Our aim in
this paper will be to write down the $N=4$ SSDYM action in a form which is
both manifestly Lorentz-invariant and supersymmetric. In some sense it is a
covariantization of the light-cone action of Siegel \cite{Siegel1,Siegel2},
obtained with the help of harmonic variables for one of the $SL(2,R)$ factors
of the Lorentz group $SO(2,2)\sim SL(2,R)_L\times SL(2,R)_R$. Harmonic
superspace
\cite{harmonics} has proved the adequate tool for manifestly supersymmetric
formulations of many supersymmetric theories. The $N=4$ SSDYM
theory is just another example in this series. It should be mentioned that
some time ago a different variational principle reproducing the self-duality
condition on (non-supersymmetric) Yang-Mills fields
has been proposed in \cite{KalSok}. It used $SU(2)$ harmonics and involved a
non-propagating Lagrange multiplier. However, according to the analysis in
ref. \cite{Markus}, it does not describe any scattering and thus cannot be
considered as a conventional field-theory action.

Self-dual Yang-Mills and supergravity have been studied from a
different point of view in a series of papers by Devchand and Ogievetsky
\cite{VI1}-\cite{VI3}. There the accent was on parametrizing all the solutions
of such self-dual equations and eventually constructing some of those solution.
In a sense, they considered a kind of a twistor transform of the self-dual
theories based on the harmonic superspace formalism. In this paper we shall
use a similar formalism, but our main purpose will
be to write down an {\it action} for the $N=4$ SSDYM theory rather
than to look for solutions to its field equations.

In section 2 we recall some basic facts about harmonic superspace. We apply the
formal rules developed for the case of $SU(2)/U(1)$ harmonics \cite{harmonics},
ignoring possible subtleties due to the non-compactness of the coset
$SL(2,R)/GL(1,R)$ in the case of Lorentz harmonics under consideration. In
section 3
we use harmonic superspace to rewrite the constraints of $N$-extended SSDYM
in the form of integrability conditions. It then becomes possible to formulate
the theory in terms of three guge connections depending on one fourth of
the original
number of Grassmann variables. This is in fact the covariantization of the
light-cone superspace used by Siegel (see also \cite{VI2}). In it the
self-duality
equations have the form of zero-curvature conditions for the three gauge
connections
(the harmonic connection and the two harmonic projections of the space-time
connection). This immediately suggests to write down an action of the
{\it Chern-Simons type}. The $GL(1,R)$ weight (closely related to the physical
dimension of the fields) of the Chern-Simons form only matches that of the
superspace measure in the maximal case of $N=4$ SSDYM. So, this action only
makes sense for $N=4$, although the same constraints and the same
Chern-Simons form can be written down for any value $0\leq N \leq 4$.

\section{Harmonic superspace with signature $(2,2)$}

The space with signature $(2,2)$ can be parametrized by coordinates
$x^{\alpha\alpha'}$, where $\alpha$ and $\alpha'$ are spinor indices of
$SL(2,R)_L$ and $SL(2,R)_R$ (the Lorentz group is $SO(2,2) \sim
SL(2,R)_L\times SL(2,R)_R$). The $N$-extended
superspace has coordinates
\begin{equation}\label{1}
x^{\alpha\alpha'}\ ,  \ \theta^\alpha_a\ ,  \  \theta^{\alpha'a} \ ,
\end{equation}
where $a$ are co- or contravariant indices of the automorphism group
$GL(N, R)$.  In it one can realize $N$-extended
supersymmetry in the following way
\begin{equation}\label{1'}
\delta x^{\alpha\alpha'} = -{1\over 2}(\epsilon^\alpha_a\theta^{\alpha'a}
+ \epsilon^{\alpha'a}\theta^\alpha_a)\ ,
\ \ \delta\theta^\alpha_a = \epsilon^\alpha_a\ ,  \ \ \delta\theta^{\alpha'a}
= \epsilon^{\alpha'a} \ .   \end{equation}
The corresponding algebra of supercovariant derivatives is
\begin{equation}\label{0}
\{D^a_\alpha,  D^b_\beta\} =0\ ,  \quad \{D_{\alpha'a},  D_{\beta'b}\}=0\ ,
\quad \{D^a_\alpha,  D_{\beta'b}\} = \delta^a_b \partial_{\alpha\beta'} \ .
\end{equation}

We choose to ``harmonize" one half of the
Lorentz group,  e.g.,  the factor $SL(2,R)_R$.  To this end we introduce real
harmonic variables $u^{\pm \alpha'}$ defined as two $SL(2,R)_R$ spinors forming
an $SL(2,R)_R$ matrix:
\begin{equation}\label{2}
u^{\pm \alpha'} \in SL(2,R)_R \ : \ \ \ u^{+\alpha'} u^-_{\alpha'} = 1
\end{equation}
(raising and lowering the $SL(2,R)_R$ spinor indices is done with the
$\epsilon$ tensor).
The index $\pm$ refers to the weight of these variables with respect to
transformations of $GL(1,R)_R \subset SL(2,R)_R$.  Thus, the harmonic
variables defined in this way should describe the non-compact coset
$SL(2,R)_R/GL(1,R)_R$.  However, we are going to apply to them the formal
rules of harmonic calculus on the compact coset $SU(2)/U(1)$
\cite{harmonics}.  In a certain sense,  this
corresponds to making a Wick rotation from signature (2,2) to (4,0).  It
is beyond the scope of this paper to give a rigorous justification of this
approach.  Nevertheless,  the formal rules will allow us to write down a
superspace action which has the correct component content.

Here we give a short summary of the rules of harmonic calculus which we are
going to use.  Harmonic functions are defined by their harmonic expansion
\begin{equation}\label{3}
f^{(q)}(u) = \sum^\infty_{n=0} f^{\alpha'_1 \ldots \alpha'_{2n+q}}
u^+_{(\alpha'_1} \ldots u^+_{\alpha'_{n+q}} u^-_{\alpha'_{n+q+1}}
\ldots u^-_{\alpha'_{2n+q})} \ .
\end{equation}
By definition, they are homogeneous under the action of $GL(1,R)_R$, i.e.,
they carry a certain weight $q$ (in (\ref{3}) $q\geq 0$).   From (\ref{3})
it is clear that the harmonic functions are collections of infinitely many
irreducible representations of $SL(2,R)_R$ (multispinors).

The principal differential operator compatible with the defining constraint
(\ref{2}) is the harmonic derivative
\begin{equation}\label{4}
\partial^{++}= u^{+\alpha'}{\partial\over\partial u^{-\alpha'}} \ : \ \ \
\partial^{++}u^{+\alpha'} =0\ ,\  \ \partial^{++} u^{-\alpha'} = u^{+\alpha'}
\ . \end{equation}
The other harmonic derivative on the two-dimensional coset
$SL(2,R)_R/GL(1,R)_R$ is $\partial^{--}$ ($\partial^{--}u^{-\alpha'} =0\ , \
\partial^{--} u^{+\alpha'} =
u^{-\alpha'}$),  but we shall never make use of it. \footnote{There exists
yet another derivative compatible with (\ref{2}), $\partial^0$
($\partial^0 u^\pm_{\alpha'} =
\pm u^\pm_{\alpha'}$). As follows from (\ref{3}), it just counts the
$GL(1,R)_R$ weight of the harmonic functions, $\partial^0f^{(q)}(u) = q
f^{(q)}(u)$.}

A direct consequence of the above definitions is the following lemma:
\begin{equation}\label{5}\partial^{++}f^{(q)}(u) = 0 \ \ \Rightarrow \
\ \left\{ \begin{array}{ll} f^{(q)}(u) = 0\ ,  \ \ q<0 \\ \\ f^{(q)}(u) =
f^{\alpha'_1 \ldots \alpha'_{q}} u^+_{(\alpha'_1} \ldots u^+_{\alpha'_{q})}
\ , \ \ q\geq 0 \end{array} \right.  \ . \end{equation}

Finally,  harmonic integration amounts to projecting out the singlet part of
a weightless integrand,  according to the formal rule
\begin{equation}\label{6}
\int du\; f^{(q)}(u) =  \left\{ \begin{array}{ll}
0,  \ q \neq 0 \\ f_{singlet} ,  \ q= 0 \end{array} \right.
\ .
\end{equation}
This integration rule is designed to give a Lorentz-invariant result. It
is compatible with integration by parts for the harmonic
derivative $\partial^{++}$.

With the help of the above harmonic variables we can define Lorentz-covariant
$GL(1,R)_R$ projections of the supercovariant derivatives from (\ref{0}),
\begin{equation}\label{7}
D^+_a = u^{+\alpha'} D_{\alpha'a},  \ \ \partial^+_\alpha = u^{+\alpha'}
\partial_{\alpha\alpha'}\ .
\end{equation}
Together with the harmonic derivative $\partial^{++}$ they form an algebra
equivalent to the original one (\ref{0}):
\begin{eqnarray}
&&\{D^a_\alpha,  D^b_\beta\} =[D^a_\alpha,  \partial^+_{\beta}] =0,  \
\{D^a_\alpha,  D^+_{b}\} = \delta^a_b \partial^+_{\alpha} \ ; \label{8} \\
&&\{D^+_{a},  D^+_{b}\}= [D^+_a,  \partial^+_{\beta}] =
[\partial^+_\alpha,  \partial^+_{\beta}] = 0 \ ;
\label{9} \\
&&[\partial^{++}, D^a_\alpha] =  [\partial^{++}, D^+_a] =
[\partial^{++}, \partial^+_{\alpha}] = 0 \ .  \label{10}
\end{eqnarray}
To see the equivalence it is sufficient to apply the lemma (\ref{5}) to
the commutation relations (\ref{10}) and thus restore the unprojected
derivatives from (\ref{7}).  When removing the harmonics $u^{+\alpha'},
u^{+\beta'}$ from a relation like, e.g., $\{D^+_{a}, D^+_{b}\}=0$ we could,
in principle, obtain terms proportional to
$\epsilon_{\alpha'\beta'}$ in the right-hand side.  However, the Lorentz index
structure and the dimensions of the available superspace operators do not
allow this (except for possible central charge terms, which we do not consider)
and we reconstruct the original algebra (\ref{0}).

The structure of the algebra (\ref{8}),  (\ref{9}) suggests several new
realizations of the $N$-extended supersymmetry algebra in subspaces of the
harmonic superspace involving only part of the Grassmann variables
$\theta$. One of them is the chiral superspace which does not contain the
variables $\theta^\alpha_a$. It is characterized by the coordinate shift
\begin{equation}\label{CB}
\mbox{Chiral basis:} \ \ \ x^{\alpha\alpha'} \ \rightarrow \ x^{\alpha\alpha'}
-{1\over 2}
\theta^\alpha_a \theta^{\alpha' a} \ .
\end{equation}
In addition, we shall regard the
following $GL(1,R)_R$ harmonic \footnote{Of course, chiral
superspace can be defined without harmonic variables (as follows from the
algebra (\ref{0})). However, the latter will be needed below
for the purpose of writing down an action for the $N=4$ SSDYM theory.}
projections as independent variables:
$$
x^{\pm \alpha} = u^\pm_{\alpha'}x^{\alpha\alpha'} \ , \qquad \theta^{\pm
a} = u^\pm_{\alpha'} \theta^{a\alpha'}\ .
$$
In this basis
the covariant derivatives from (\ref{8})-(\ref{10}) become
\begin{equation}\label{11'}
D^a_\alpha = \partial^a_\alpha  \ , \quad
D^+_a = \partial^+_a + \theta^\alpha_a \partial^+_\alpha \ , \quad D^{++}
= \partial^{++} + \theta^{+a}
\partial^+_a + x^{+\alpha}\partial^+_\alpha \ .
\end{equation}
Here and in what follows we use the notation
$$
\partial^a_\alpha = {\partial\over\partial\theta^\alpha_a} \ , \qquad
\partial^\pm_a = {\partial\over\partial \theta^{\mp a}}\ , \qquad
\partial^\pm_\alpha = {\partial\over\partial x^{\mp \alpha}}\ .
$$
Note the appearance of vielbein terms in the harmonic derivative
$D^{++}$ in (\ref{11'}). So, in this basis the {\it chiral} superfields
defined by the constraint
\begin{equation}\label{chiral}
D^a_\alpha \Phi = 0  \ \ \ \Rightarrow \ \ \
\Phi=\Phi(x^{\pm\alpha},\theta^{\pm a}, u)
\end{equation}
do not depend on $\theta^\alpha_a$.
In the chiral subspace the supersymmetry transformations are
\begin{equation}\label{11''}
\delta x^{\pm \alpha} = - \epsilon^\alpha_a \theta^{\pm a} \ , \quad
\delta \theta^{\pm a} = u^\pm_{\alpha'}\epsilon^{\alpha' a} \ ,
\quad \delta u^\pm_{\alpha'} = 0 \ .
\end{equation}

Another possibility offered by the algebra (\ref{8}),  (\ref{9}) is to
eliminate the projections $\theta^{-a}$ from the supersymmetry
transformations. To this end one makes the shift
$x^{-\alpha} \ \rightarrow \ x^{-\alpha}
+ {1\over 2}
\theta^\alpha_a \theta^{- a} $, after which the spinor derivative
$D^+_a$ becomes $D^+_a = \partial/\partial
\theta^{-a}$. Then the {\it analytic} superfields defined by
\begin{equation}\label{analytic}
D^+_a \Phi = 0  \ \ \ \Rightarrow \ \ \
\Phi=\Phi(x^{\pm\alpha},\theta^{+ a}, \theta^\alpha_a, u)
 \end{equation}
 do not depend on  $\theta^{-a}$.
In this analytic subspace supersymmetry acts as follows:
\begin{equation}\label{ABS}
\delta x^{+\alpha} = -{1\over 2}(\epsilon^\alpha_a \theta^{+a}
+ \epsilon^{+a} \theta^\alpha_a) \ , \quad
\delta x^{-\alpha} = -\epsilon^{-a} \theta^\alpha_a \ , \quad
\delta \theta^{+a} = u^+_{\alpha'}\epsilon^{\alpha' a} \ ,\quad
\delta \theta^\alpha_a = \epsilon^\alpha_a \ ,
\quad \delta u^\pm_{\alpha'} = 0 \ .
 \end{equation}

 A peculiarity of the harmonic superspace under consideration
 is the existence \cite{VI1} of an even
 smaller superspace containing only $\theta^{+a}$. It is defined by
 imposing the chirality (\ref{chiral}) and analyticity (\ref{analytic})
 constraints {\it simultaneously}:
\begin{equation}\label{anchir}
D^a_\alpha\Phi = D^+_a \Phi = \partial^+_\alpha \Phi = 0 \ \ \
\Rightarrow \ \ \ \Phi=\Phi(x^{+\alpha},\theta^{+a},u) \ .
 \end{equation}
Note that the third constraint is an inevitable corollary of the first
two and of the
anticommutation relations (\ref{8}). In a
suitable superspace basis supersymmetry is
realized on $x^{+\alpha}$ and $\theta^{+a}$ only:
\begin{equation}\label{ACS}
\delta x^{+\alpha} = -\epsilon^\alpha_a \theta^{+a} \ ,
\quad \delta \theta^{+a} = u^+_{\alpha'}\epsilon^{\alpha' a}  \ ,
\quad \delta u^\pm_{\alpha'} = 0 \ .
  \end{equation}
Such superfields are automatically on shell, since
\begin{equation}\label{onshell}
\partial^+_\alpha \phi = 0 \ \
\Rightarrow \ \ \ \Box \phi = 2\partial^{+\alpha}\partial^-_\alpha
\Phi = 0 \ .
 \end{equation}

\section{Self-dual supersymmetric Yang-Mills theory }

\subsection{Superspace constraints}

$N$-extended ($0\leq N \leq 4$) supersymmetric Yang-Mills theory is
described by the algebra of the gauge-covariantized superspace derivatives
from (\ref{0}):
\begin{eqnarray}
&&\{\nabla^a_\alpha,  \nabla^b_\beta\}
=\epsilon_{\alpha\beta}\tilde\phi^{ab}\ ,  \ \ [\nabla^a_\alpha,
\nabla_{\beta\beta'}]= \epsilon_{\alpha\beta} \tilde\chi^a_{\beta'} \ ;
\label{12''} \nonumber \\
&&\{\nabla^a_\alpha,  \nabla_{\beta'b}\} =
\delta^a_b \nabla_{\alpha\beta'} \ ; \label{12'} \\
&&\{\nabla_{\alpha'a},  \nabla_{\beta'b}\}= \epsilon_{\alpha'\beta'}
\phi_{ab}\ ,  \ [\nabla_{\alpha'a}, \nabla_{\beta\beta'}] =
\epsilon_{\alpha'\beta'} \chi_{\beta a}\ ; \nonumber \\
&&[\nabla_{\alpha\alpha'},  \nabla_{\beta\beta'}] =
\epsilon_{\alpha'\beta'} F_{\alpha\beta} + \epsilon_{\alpha\beta}
F_{\alpha'\beta'}
\ , \nonumber
\end{eqnarray}
where $\tilde\phi^{ab} = - \tilde\phi^{ba}$,  $\phi^{ab} = - \phi^{ba}$ (for
$N=1$ the scalars $\phi$ drop out).  In the non-supersymmetric case $N=0$ only
the last relation in (\ref{12'}) remains. For $N=0,1, 2$ the theory is off
shell, whereas for $N=3, 4$ it is on shell.  In addition,  in the case $N=4$
one should require the two sets of 6 scalars to be related,
$\tilde\phi^{ab}={1\over 2}\epsilon^{abcd}\phi_{ab}$\ .
\footnote{This is the analog of the reality
condition on the scalars in $N=4$ SYM theory in the case of minkovskian
signature (1,3) \cite{N=4}.  In the case of signature (2,2) the scalars are
real by definition. }

Self-duality means that half of the field strengths vanish,  e.g.,  all those
appearing in (\ref{12'}) multiplied by $\epsilon_{\alpha\beta}$ (of course,
the $N=4$ relation $\tilde\phi^{ab} =
{1\over 2}\epsilon^{abcd}\phi_{ab}$ does not hold any longer).
Thus we obtain the constraints of SSDYM theory
\begin{eqnarray}
&&\{\nabla^a_\alpha,  \nabla^b_\beta\} =0,  \ \{\nabla^a_\alpha,
\nabla_{\beta'b}\} =
\delta^a_b \nabla_{\alpha\beta'}\ ,  \
\ [\nabla^a_\alpha,  \nabla_{\beta\beta'}]=0\ ; \label{12} \\
&&\{\nabla_{\alpha'a},  \nabla_{\beta'b}\}= \epsilon_{\alpha'\beta'}
\phi_{ab}\ ,  \ [\nabla_{\alpha'a}, \nabla_{\beta\beta'}] =
\epsilon_{\alpha'\beta'} \chi_{\beta a}\ , \ [\nabla_{\alpha\alpha'},
\nabla_{\beta\beta'}] = \epsilon_{\alpha'\beta'} F_{\alpha\beta}\
. \label{13} \end{eqnarray}
Since the self-duality condition $F_{\alpha'\beta'} =0$
on the Yang-Mills field is a
dynamical equation,  the constraints (\ref{12}),  (\ref{13}) now describe an
on-shell theory for any value $0\leq N \leq 4$. This supermultiplet
contains helicities from $+1$ down to $1-{N\over 2}$. Clearly, it only
becomes self-conjugate, i.e., spans all the helicities from $+1$ to $-1$ in the
maximal case $N=4$. As a consequence, in the latter case the degrees of
freedom appear in Lagrangian pairs and one is able to write down an
action for the theory. So far this action
has been presented either in component (i.e., not manifestly
supersymmetric) \cite{Siegel2,Gates}
or light-cone superspace (i.e., not manifestly Lorentz-invariant) form
\cite{Siegel1,Siegel2}. Our purpose in this paper will be to write down the
$N=4$ SSDYM action in a form which is both manifestly Lorentz-invariant and
supersymmetric. To this end we shall first relax (\ref{12}),  (\ref{13}) in
order to go off shell and then we shall find a variational principle
from which (\ref{12}), (\ref{13}) will follow as field equations.

Our first step will be to obtain a set of
(anti)commutation relations completely free from curvatures with the help
of the harmonic variables introduced in section 2. \footnote{Our treatment
of the SSDYM constraints is, up to a certain point, similar to that in
\cite{VI2}.} Defining the harmonic projections (cf. (\ref{7}))
\begin{equation}\label{16}
\nabla^+_a = u^{+\alpha'} \nabla_{\alpha'a}\ ,  \ \
\nabla^+_\alpha = u^{+\alpha'}
\nabla_{\alpha\alpha'}\ ,  \end{equation}
we obtain from (\ref{12}), (\ref{13})
\begin{eqnarray}
&&\{\nabla^a_\alpha,  \nabla^b_\beta\} =0\ , \label{141} \\
&&\{\nabla^a_\alpha,
\nabla^+_{b}\} = \delta^a_b \nabla^+_{\alpha}\ , \label{142} \\
&&[\nabla^a_\alpha,  \nabla^+_{\beta}]=0\ , \label{143} \\
&&\{\nabla^+_{a},  \nabla^+_{b}\}= 0\ , \label{144} \\
&&[\nabla^+_{a}, \nabla^+_{\beta}] = 0\ , \label{145} \\
&&[\nabla^+_{\alpha},  \nabla^+_{\beta}] = 0 \ . \label{146}
\end{eqnarray}
In fact,  these constraints are equivalent to the initial set (\ref{12}),
(\ref{13}).  To see this one takes into account the linear
harmonic dependence of the projected covariant derivatives (\ref{16}) and
then pulls out the harmonics $u^+$ from the relations (\ref{142})-(\ref{146}).
In doing so the terms proportional to $\epsilon_{\alpha'\beta'}$
appear in the right-hand side of eqs.  (\ref{13}).
The information contained in (\ref{16})
can also be encoded in the form of commutation relations with the harmonic
derivative $\partial^{++}$ (cf.  (\ref{10}) and recall (\ref{5}))
\begin{equation}\label{17}
[\partial^{++}, \nabla^a_\alpha] =
[\partial^{++}, \nabla^+_a] =
[\partial^{++}, \nabla^+_{\alpha}] = 0 \ .
\end{equation}
This means that we first assume that the gauge connections $A^a_\alpha,
A^+_a, A^+_\alpha$ are arbitrary functions of the harmonic variables
$u^\pm_{\alpha'}$. Then the r\^ole of the constraints (\ref{17}) is to
reduce this dependence to a trivial one. In fact, we can go a step
further and start from a framework in which
not only the gauge connections but also the gauge group parameters
have an arbitrary dependence of the harmonic variables.
This implies that the harmonic derivative
$\partial^{++}$ is covariantized as well,
$$
\nabla^{++} = \partial^{++} + A^{++}(x,\theta,u) \ .
$$
Then eqs. (\ref{17}) are replaced by covariant ones,
\begin{eqnarray}
&&[\nabla^{++}, \nabla^a_\alpha] =  0 \ , \label{011} \\
&&[\nabla^{++}, \nabla^+_a] =  0 \ , \label{012} \\
&&[\nabla^{++}, \nabla^+_{\alpha}] = 0 \ . \label{013}
\end{eqnarray}
In order to go back to the frame in which the harmonic dependence is
trivial it is sufficient to  eliminate the newly introduced harmonic
connection $A^{++}(x,\theta,u)$ by a suitable harmonic-dependent gauge
transformation:
$$
A^{++}(x,\theta,u) = e^{-\Lambda(x,\theta,u)} \partial^{++}
e^{\Lambda(x,\theta,u)} \ .
$$
This is always possible, since there is only one such connection (no
integrability conditions).
Then we recover the original constraints (\ref{17}), from which we
deduce the trivial harmonic dependence of the remaining connections
$A^a_\alpha,A^+_a, A^+_\alpha$.

For our purposes it will be preferable to stay in the frame
with non-trivial harmonic dependence of the gauge
objects. Even so, the constraints (\ref{141})-(\ref{146}) and
(\ref{011})-(\ref{013}) still allow us to choose alternative special
gauge frames. One possibility typical for other
harmonic gauge theories (see
\cite{harmonics,hartw,N=3})
would be to use the zero-curvature constraint (\ref{144}) and gauge
away the connections $A^+_a$ (``analytic frame''). In such a frame the
notion of an analytic ($\theta^{-a}$-independent) superfield (\ref{analytic})
is preserved.
However, we do not find it useful in the present context. Instead, we can
choose a chiral gauge frame in which the connection
$A^a_\alpha$ vanishes (its existence is guaranteed by the zero-curvature
condition (\ref{141})). Although this could be done even before introducing
harmonic variables, the relevance of the latter will become clear shortly.
So, using the chiral basis (\ref{CB}) for the spinor
derivatives, we can trivialize the covariant derivative $\nabla^a_\alpha$:
\begin{equation}\label{00}
\mbox{Chiral gauge:} \ \ \ \ A^a_\alpha = 0 \ \ \rightarrow \ \
\nabla^a_\alpha = \partial^a_\alpha \ .
\end{equation}
Note that in this new gauge frame (\ref{00}) the gauge parameters are
chiral (i.e., independent of $\theta^{\alpha}_a$) but still harmonic
dependent, $\Lambda(x, \theta^{\pm a}, u) $.
Further, from (\ref{143}), (\ref{00}) and (\ref{011}) we find
\begin{equation}\label{02}
\partial^a_\alpha A^+_\beta = \partial^a_\alpha A^{++} = 0 \ \ \Rightarrow
\ \ A^+_\alpha = A^+_\alpha (x, \theta^{\pm a}, u)
\ , \ \ \ A^{++} = A^{++} (x, \theta^{\pm a}, u) \ .
\end{equation}
Eq. (\ref{142}) then has the general solution
\begin{equation}\label{03}
A^+_a = a^+_a (x, \theta^{\pm a}, u) + \theta^\alpha_a A^+_\alpha (x,
\theta^{\pm a}, u) \ .
\end{equation}
Substituting (\ref{03}) into (\ref{144}), using (\ref{11'}) and
collecting the terms with 0, 1 and 2 $\theta^\alpha_a\;$,
we obtain the following constraints:
\begin{eqnarray}
&&\partial^+_a a^+_b + \partial^+_b a^+_a + \{a^+_a,a^+_b\} = 0 \ ,
\label{041} \\
&&\partial^+_a A^+_\beta - \partial^+_\beta a^+_a + [a^+_a, A^+_\beta]  = 0 \ ,
\label{042} \\
&&\partial^+_\alpha A^+_\beta - \partial^+_\beta A^+_\alpha + [A^+_\alpha,
A^+_\beta] = 0 \ . \label{043}
\end{eqnarray}
The first of them implies that the part
$a^+_a(x,\theta^{\pm a}, u)$ of $A^+_a$ (\ref{03}) is pure gauge and can be
gauged away by a suitable gauge transformation
\begin{equation}\label{gt}
\delta A^+_a = \partial^+_a\Lambda + [A^+_a,\Lambda]  \end{equation}
with a chiral parameter $\Lambda$.
 From now on it will be convenient to work in the
\begin{equation}\label{05}
\mbox{semianalytic gauge:} \ \ \ \ a^+_a (x, \theta^{\pm a}, u) = 0 \ .
\end{equation}
Then from (\ref{042}) we find
\begin{equation}\label{06}
\partial^+_a A^+_\beta =  0 \ \ \Rightarrow
\ \ A^+_\alpha = A^+_\alpha (x, \theta^{+ a}, u) \ .
\end{equation}
The next step is to insert all the above results in eq. (\ref{012}). The
$\theta^\alpha_a$-independent term gives
\begin{equation}\label{07}
\partial^+_a A^{++} =  0 \ \ \Rightarrow
\ \ A^{++} = A^{++} (x, \theta^{+ a}, u)
\end{equation}
and the term linear in $\theta^\alpha_a$ yields the constraint
\begin{equation}\label{08}
\partial^+_\alpha A^{++} - D^{++} A^+_\alpha + [A^+_\alpha,
A^{++}] = 0 \ .
\end{equation}
Among the remaining constraints only that on the connections $A^+_\alpha$
(\ref{043}) is independent, eqs. (\ref{145}), (\ref{146}) and (\ref{013})
then follow.

Comparing the harmonic treatment of the
SSDYM constraints given here with the more traditional approach to harmonic
gauge theories in refs. \cite{harmonics,hartw,N=3}, we see that here we
gauge away the spinor connection $A^+_a$ only partially
(\ref{05}) (the remaining part of it is related to the vector connection
$A^+_\alpha$, see (\ref{03})). At the same time, the other spinor connection
$A^a_\alpha$ is
fully gauged away (chiral gauge (\ref{00})). This mixed chiral-semianalytic
gauge explains why we needed to keep a non-trivial harmonic dependence when
introducing the chiral gauge (\ref{00}). In fact, we could
go one more step further and fix a fully analytic gauge in which
the entire spinor connection $A^+_a$ (or, equivalently, the vector
connection $A^+_\alpha$, see (\ref{03})) is gauged away. This is permitted
by the zero-curvature condition (\ref{043}).
In this case we would obtain a twistor transform of the on-shell SSDYM fields
(see the discussion around eq. (\ref{ssg})). However, for the purpose of
writing down an action we should keep the set of three gauge
connections $A^+_\alpha$, $A^{++}$ which are functions of only one fourth of
the
Grassmann variables:
\begin{equation}\label{09}
A^+_\alpha = A^+_\alpha (x,\theta^+,u) \ , \qquad A^{++} = A^{++}
(x,\theta^+,u) \ .
\end{equation}
These connections undergo gauge transformation
\begin{equation}\label{21} \delta A^+_{\alpha} = \partial^+_\alpha \Lambda
+ [A^+_{\alpha}, \Lambda]\ ,  \ \ \ \delta A^{++} = D^{++} \Lambda
+ [A^{++}, \Lambda]\ , \ \ \ \Lambda=\Lambda(x,\theta^+,u) \ ,
\end{equation}
which are compatible with the chiral-semianalytic gauge (\ref{00}),
(\ref{05}).
The connections are put on shell by the three zero-curvature conditions:
\begin{eqnarray}
&&\partial^+_\alpha A^{++} - D^{++} A^+_\alpha + [A^+_\alpha,
A^{++}] = 0 \ , \label{0102}  \\
&&\partial^+_\alpha A^+_\beta - \partial^+_\beta A^+_\alpha + [A^+_\alpha,
A^+_\beta] = 0 \ . \label{0101}
\end{eqnarray}
All this represents an equivalent reformulation of the $N$-extended SSDYM
theory and will serve as the basis for our action in the case $N=4$.
Before addressing the issue of the action, we would like to make a
number of comments.

\subsection{Supersymmetry transformations}

The gauge connections (\ref{09}) do not transform as superfields under the
right-handed part (parameters $\epsilon^\alpha_a$) of the supersymmetry
algebra (\ref{11''}). Indeed, $A^+_a$ in (\ref{03}) is a
supercovariant object but its component $a^+_a$ from the $\theta^\alpha_a$
expansion is not,
\begin{equation}\label{021}
\delta a^+_a = (\epsilon^\beta_b\theta^{+b} \partial^-_\beta
+\epsilon^\beta_b\theta^{-b} \partial^+_\beta
- \epsilon^{+b}\partial^-_b)a^+_a   -\epsilon^\alpha_a A^+_\alpha \ .
  \end{equation}
Here we have explicitly written out the supertranslation terms.
Earlier we fixed the gauge (\ref{05}) which is violated by the inhomogeneous
term in (\ref{021}).
In order to correct this we have to accompany the supersymmetry transformation
by a compensating gauge transformation (\ref{gt}) with parameter
\begin{equation}\label{022}
\Lambda = (\theta^{-a}\epsilon^\alpha_a) A^+_\alpha(x,\theta^+,u) \ .
\end{equation}
This gauge transformation affects the connections $A^+_\alpha$, $A^{++}$
as well, so their supersymmetry transformation laws are modified:
\begin{eqnarray}
\delta A^+_\alpha &=& (\epsilon^\beta_b\theta^{+b} \partial^-_\beta
- \epsilon^{+b}\partial^-_b)A^+_\alpha + (\theta^{-b}\epsilon^\beta_b)
(\partial^+_\alpha A^+_\beta - \partial^+_\beta A^+_\alpha + [A^+_\alpha,
A^+_\beta]) \ , \label{0231}\\
\delta A^{++} &=& (\epsilon^\beta_b\theta^{+b} \partial^-_\beta
- \epsilon^{+b}\partial^-_b)A^{++}
 + (\theta^{-b}\epsilon^\beta_b)
(D^{++} A^+_\beta - \partial^+_\beta A^{++} + [A^{++}, A^+_\beta])
\nonumber \\
&+& (\theta^{+b}\epsilon^\beta_b) A^+_\beta\ . \label{0232}
\end{eqnarray}
The terms containing
$\theta^-$ are proportional to the constraints (\ref{0102}),
(\ref{0101}), so they drop out. Remarkably, the connections
$A^+_\alpha(x,\theta^+,u)$, $A^{++}(x,\theta^+,u)$ transform as if the
supersymmetry transformations (\ref{11''}) did not involve $\theta^-$ at all:
\begin{eqnarray}
\delta A^+_\alpha &=& (\epsilon^\beta_b\theta^{+b} \partial^-_\beta
- \epsilon^{+b}\partial^-_b)A^+_\alpha  \ , \label{0241}\\
\delta A^{++} &=& (\epsilon^\beta_b\theta^{+b} \partial^-_\beta
- \epsilon^{+b}\partial^-_b)A^{++}
 + (\theta^{+b}\epsilon^\beta_b) A^+_\beta\ . \label{0242}
\end{eqnarray}
It must be stressed that these objects should not be confused with
the chiral-analytic superfields defined by eq. (\ref{anchir}). The latter do
not
depend on $x^{-\alpha}$ and are thus automatically on shell. For our purposes
it is essential that the superfields $A^+_\alpha$, $A^{++}$ can
exist off shell too. Therefore we shall never require
$\partial^+_\alpha A = 0$.

Thus, eqs. (\ref{0241}), (\ref{0242}) are the transformation laws of
``semicovariant" superfields. Another way to say this is to point out that
commuting two such supersymmetry transformations
one obtains the required translations in the direction
$x^{-\alpha}$ only with the help of the constraint (\ref{0102}) and of a
compensating gauge transformation (\ref{gt}) with parameter
$\Lambda = (\epsilon^{-a}_1 \epsilon^\alpha_{2a} -
\epsilon^{-a}_2 \epsilon^\alpha_{1a}) A^+_\alpha$ (this follows from
(\ref{022}) as well). From this point of view the
harmonic derivative $D^{++}$ (\ref{11'}) is not supercovariant either,
i.e., it does not commute with the translation part of
the transformations (\ref{0241}), (\ref{0242}):
\begin{equation}\label{025}
[D^{++}\ , \ \epsilon^\beta_b\theta^{+b} \partial^-_\beta
- \epsilon^{+b}\partial^-_b] = -\epsilon^\beta_b\theta^{+b} \partial^+_\beta
+ \epsilon^{+b}\partial^+_b \ .
 \end{equation}

The supersymmetry transformation rules (\ref{0241}), (\ref{0242}),
in which $\theta^-$ never appears, were obtained using
the on-shell constraints (\ref{0102}), (\ref{0101}). When
writing down the $N=4$ action in
section 4 we shall treat the connections
$A^+_\alpha(x,\theta^+,u)$, $A^{++}(x,\theta^+,u)$ as {\it unconstrained}
objects. Nevertheless, we shall apply the same supersymmetry rules to them.
What is important in
this context is to make sure that the left-hand sides of the constraints
still form a supersymmetric set, i.e., that they transform into each other.
Indeed, this is easy to check
using the transformation laws (\ref{0241}), (\ref{0242}), (\ref{025}).

\subsection{Components}

Now we would like to give a direct demonstration that the constraints
(\ref{0102}),
(\ref{0101}) do indeed describe $N$-extended SSDYM theory. To this end we shall
exhibit the component content of the gauge superfields
$A^+_\alpha(x,\theta^+,u)$, $A^{++}(x,\theta^+,u)$. Let us first consider the
simplest case $N=1$. The harmonic
connection has a very short Grassmann expansion:
\begin{equation}\label{-22}
A^{++} = a^{++}(x, u) + \theta^{+}\sigma^+(x, u) \ .
\end{equation}
The fields in (\ref{-22}) are harmonic, i.e., they contain infinitely many
ordinary fields (recall (\ref{3})). However, we still have the gauge
transformations
(\ref{21}) with parameter
\begin{equation}\label{-22'}
\Lambda = \lambda(x, u) + \theta^{+}\rho^-(x, u)\ .  \end{equation}
Let us compare the harmonic expansions (\ref{3}) of the bosonic components in
(\ref{-22}) and (\ref{-22'}):
\begin{eqnarray}
&&a^{++}(x, u) = u^+_{(\alpha'}u^+_{\beta')} a^{\alpha'\beta'}(x) +
u^+_{(\alpha'}u^+_{\beta'} u^+_{\gamma'}u^-_{\delta')}
a^{\alpha'\beta'\gamma'\delta'}(x) + \ldots \ , \nonumber \\
&&\lambda(x, u) = \lambda(x) + u^+_{(\alpha'}u^-_{\beta')}
\lambda^{\alpha'\beta'}(x) +
u^+_{(\alpha'}u^+_{\beta'} u^-_{\gamma'}u^-_{\delta')}
\lambda^{\alpha'\beta'\gamma'\delta'}(x) + \ldots \ . \nonumber
\end{eqnarray}
Clearly, the parameter $\lambda(x, u)$ contains enough components to completely
gauge away the harmonic field $a^{++}(x, u)$ (note that the singlet part
$\lambda(x)$ in $\lambda(x, u)$ is not used in the process; it remains
non-fixed and plays the r\^ole of the ordinary gauge parameter). Similarly, the
parameter $\rho^-(x, u)$ can gauge away the entire field $\sigma^+(x, u)$.
Thus, we arrive at the following
\begin{equation}\label{-23}
\mbox{$N=1$ Wess-Zumino gauge:}\qquad A^{++}= 0\ .
\end{equation}

The other gauge connection has the expansion
\begin{equation}\label{-25}
A^{+}_\alpha = {\cal A}^{+}_\alpha(x, u) + \theta^{+}
\chi_{\alpha }(x, u) \ .
\end{equation}
The harmonic dependence in it can be eliminated by using the
constraint (\ref{0102}).  Substituting (\ref{-23}) and (\ref{-25})
into (\ref{0102}), we obtain $D^{++}A^+_\alpha = 0$, from where follow the
harmonic equations
\begin{eqnarray}
&&\partial^{++}{\cal A}^+_\alpha(x, u) = 0 \ \ \Rightarrow
{\cal A}^+_\alpha(x, u) = u^{+\alpha'}{\cal A}_{\alpha\alpha'}(x) \ , \nonumber
\\
&&\partial^{++}\chi_{\alpha }(x, u) = 0 \ \ \Rightarrow
\chi_\alpha(x, u) = \chi_\alpha(x) \ . \label{hars}
\end{eqnarray}
Then, inserting (\ref{hars}) into the remaining constraint (\ref{0101}),
we obtain the self-duality equation for the Yang-Mills field
${\cal A}_{\alpha\alpha'}(x)$ and the Dirac equation for the chiral spinor
$\chi_\alpha(x)$\ :
\begin{equation}\label{em}
{F}^{\alpha'\beta'} = \partial^{\alpha(\alpha'}
{\cal A}_\alpha^{\beta')} + {\cal A}^{\alpha(\alpha'}{\cal A}^{\beta')}_\alpha
= 0 \ ,
\qquad \nabla^{\alpha\alpha'}\chi_\alpha = 0 \ ,
\end{equation}
where $\nabla_{\alpha\alpha'} = \partial_{\alpha\alpha'} +
[{\cal A}_{\alpha\alpha'}\ , \ ]$ denotes the usual Yang-Mills covariant
derivative. This is precisely the content of the $N=1$ SSDYM multiplet.

It is not hard to find out the supersymmetry transformation laws of the
component fields ${\cal A}_{\alpha\alpha'}$, $\chi_\alpha$. To this end we
note that in order for the supersymmetry transformation (\ref{0242}) not to
violate the Wess-Zumino gauge (\ref{-23}), we have to make a compensating
gauge transformation (\ref{21}) with parameter $\Lambda = (\theta^+
\epsilon^\alpha){\cal A}^-_\alpha$
(where ${\cal A}^-_\alpha = u^{-\alpha'}{\cal A}_{\alpha\alpha'}(x)$). Then
the combination of (\ref{0241}) with this gauge transformation gives
$$
\delta A^+_\alpha = -\epsilon^+\chi_\alpha - \theta^+\epsilon^\beta
(\partial^-_\beta {\cal A}^+_\alpha + \partial^+_\alpha {\cal A}^-_\beta
+[{\cal A}^-_\alpha, {\cal A}^+_\beta]) \ .
$$
This, together with the field equation for ${\cal A}$ (\ref{em}) leads to
the transformation laws
$$
\delta {\cal A}_{\alpha\alpha'} = -\epsilon_{\alpha'} \chi_\alpha \ , \qquad
\delta \chi_\alpha = -{1\over 2} \epsilon^\beta {F}_{\alpha\beta} \ ,
$$
where ${F}_{\alpha\beta}$ is the self-dual part of the Yang-Mills curvature.

The maximal case $N=4$ follows the same pattern. There one can fix the
\begin{equation}\label{23}
\mbox{$N=4$ WZ gauge:}\qquad A^{++}= (\theta^+)^{2ab}\phi_{ab}(x) +
(\theta^+)^3_a
u^{-\alpha'}\chi^{a}_{\alpha'}(x) + (\theta^+)^4 u^{-\alpha'}u^{-\beta'}
G_{\alpha'\beta'}(x)\ ,
\end{equation}
where
$$
(\theta^+)^{2ab} = {1\over 2!}\theta^{+a}\theta^{+b},  \ (\theta^+)^3_a  =
{1\over 3!} \epsilon_{abcd}\theta^{+b}\theta^{+c}\theta^{+d},  \
(\theta^+)^4 = {1\over 4!}
\epsilon_{abcd}\theta^{+a}\theta^{+b}\theta^{+c}\theta^{+d}
\ .  $$
The other gauge connection has the expansion
\begin{equation}\label{25}
A^{+}_\alpha = {\cal A}^{+}_\alpha(x, u) + \theta^{+a}
\chi_{\alpha a}(x, u) +
(\theta^+)^{2ab}B^-_{\alpha ab}(x, u) + (\theta^+)^3_a \tau^{--a}_\alpha(x, u)
+ (\theta^+)^4 C^{-3}_\alpha(x, u)\ .
\end{equation}
With the help of the constraint (\ref{0102}) one eliminates
the harmonic dependence in (\ref{25}),
\begin{eqnarray}
&& {\cal A}^+_\alpha(x, u) = u^{+\alpha'}{\cal A}_{\alpha\alpha'}(x)\ ,
\qquad
\chi_{\alpha a}(x, u) = \chi_{\alpha a}(x)\ ,  \nonumber \\
&& B^-_{\alpha ab}(x, u) = u^{-\alpha'}\nabla_{\alpha\alpha'}\phi_{ab}(x)\ ,
\qquad \tau^{--a}_\alpha(x, u) = {1\over 2}u^{-\alpha'}u^{-\beta'}
\nabla_{\alpha\alpha'}\chi^a_{\beta'}(x)\ , \nonumber \\
&& C^{-3}_\alpha(x, u) = {1\over 3}u^{-\alpha'}u^{-\beta'}u^{-\gamma'}
\nabla_{\alpha\alpha'} G_{\beta'\gamma'}(x) \ .  \label{59}
\end{eqnarray}
In addition,  the constraint (\ref{0101}) implies that the physical fields
$\phi_{ab}\ ,\ \chi_{\alpha a}\ , \ \chi^{a}_{\alpha'}$ satisfy their
equations of motion. The fields
${\cal A}_{\alpha\alpha'}$ and $G_{\alpha'\beta'}$ describe a self-dual and
an anti-self-dual gauge fields,  respectively.  Thus we find the complete
content of the $N=4$ SSDYM multiplet,  as given in
\cite{Siegel1,Siegel2,Gates}. Following the $N=1$ example, it is not hard to
also derive the supersymmetry transformation laws from \cite{Siegel2,Gates}.

Here we would like to comment on another possibility to fix the gauge.
Above we showed that a non-supersymmetric off-shell gauge of the Wess-Zumino
type is
necessary in order to obtain the standard components of the theory. However,
on shell there exists an alternative, manifestly supersymmetric gauge. Indeed,
the
constraint (\ref{0101}) tells us that $A^+_\alpha$ is {\it pure gauge
on shell}. If we fix the
\begin{equation}\label{ssg}
\mbox{supersymmetric on-shell gauge}: \ \ \
A^+_\alpha = 0 \ ,
\end{equation}
the remaining constraint (\ref{0102}) becomes simply
\begin{equation}\label{trivial}
\partial^+_\alpha A^{++}=0 \ .
\end{equation}
It means that the components of $A^{++}$ are harmonic fields
$\phi(x^{+\alpha}, u)$ independent of $x^{-\alpha}$. As explained in
(\ref{onshell}), such fields are automatically on shell.
In fact,  what we encounter here are {\it twistor-type solutions} of
massless equations of motion \cite{Penrose}.
\footnote{We should repeat here that
our harmonic variables are defined on the non-compact coset
$SL(2,R)_R/GL(1,R)_R$.  Therefore the analogy with the standard twistor
approach \cite{Penrose},
based on the compact coset $S^2\sim SU(2)/U(1)$, remains formal. } Thus,
we can say that the connection $A^{++}$
now contains the set of twistor transforms of all the fields of the
$N$-extended
SSDYM multiplet. In other words, the solutions to the SSDYM equations
are encoded in a single on-shell superfield $A^{++}$. The situation
here closely resembles the harmonic version of the twistor transform of the
ordinary ($N=0$) SDYM equations, where all the self-dual solutions (e.g.,
instantons) are parametrized by a single object
$V^{++}(x^+_\alpha, u^\pm_{\alpha'})$
\cite{hartw}. This line of study of the SSDYM system has been proposed and
pursued in \cite{VI1,VI2}. In the present paper we are interested in an {\it
action} for the $N=4$ theory, therefore the on-shell gauge (\ref{ssg}) will
not be implemented.

\subsection{Action for $N=4$ SSDYM} \label{action}

Up to now the whole discussion
applied equally well to all values $0\leq N \leq 4$.  The unique features of
the case $N=4$ only become important when one tries to write down an action.
At the component level this is manifested in the fact that the $N=4$ multiplet
contains all the helicities from $+1$ (described by the self-dual field
${\cal A}$) down to $-1$ (the field $G$).  It is precisely the field
$G_{\alpha'
\beta'}$ which can serve as a (propagating!) Lagrange multiplier for the
self-duality condition on ${\cal A}_{\alpha\alpha'}$. Similarly, the spinor
fields $\chi^a_{\alpha'}$ and $\chi_{\alpha a}$ form a Lagrangian pair. The
form of this action first given in \cite{Siegel2,Gates} is
\begin{equation}
S = \int d^4x \; \mbox{Tr}\;\left\{ -{1\over 8} \nabla^{\alpha\alpha'}\phi^{ab}
\nabla_{\alpha\alpha'} \phi_{ ab} +{1\over 2} \chi^a_{\alpha'}
\nabla^{\alpha\alpha'}
\chi_{\alpha a} + {1\over 6} G^{\alpha'\beta'} F_{\alpha'\beta'}
 - \phi^{ab} \chi^\alpha_a \chi_{\alpha b}\right\} \ .  \label{33}
 \end{equation}

The purpose of the harmonic superspace formalism developed above was to
write down an {\it action for $N=4$ SSDYM with manifest Lorentz
invariance and supersymmetry}.  So far we have rewritten the on-shell
constraints of the theory in the equivalent form (\ref{0102}),  (\ref{0101}).
Now
we want to obtain these dynamical equations from a variational principle.
Eqs. (\ref{0102}), (\ref{0101}) have the form of vanishing curvature
conditions.
Note also the important fact that we have three gauge connections
$A^{++}, A^+_\alpha$ and,  correspondingly,  three curvatures made out of them.
All this suggests to write down the {\it Chern-Simons form}
\begin{equation}\label{CS}
L^{+4}(x,\theta^+,u) =  \mbox{Tr}\; (A^{++}\partial^{+\alpha}A^+_\alpha-
{1\over 2} A^{+\alpha}D^{++} A^+_\alpha + A^{++} A^{+\alpha} A^+_\alpha) \ .
 \end{equation}
Since the connections in (\ref{CS}) are not covariant superfields, we should
find out how the Chern-Simons form transforms under supersymmetry. Using
(\ref{0241}), (\ref{0242}), (\ref{025}) it is easy to check that
\begin{equation}\label{CStr}
\delta L^{+4} = (\epsilon^\beta_b\theta^{+b} \partial^-_\beta
- \epsilon^{+b}\partial^-_b) L^{+4} +
{1\over 2} (\theta^{+b}\epsilon^\beta_b) \partial^{+\alpha}
\mbox{Tr} (A^+_\alpha A^+_\beta) \ ,
\end{equation}
i.e.,  it transforms into a total derivative with respect to the variables
$x^{\pm \alpha}$ and $\theta^{+a}$. Similarly, the gauge transformations
(\ref{21})
give
\begin{equation}\label{CSgtr}
\delta L^{+4} = \partial^{+\alpha} \mbox{Tr} (D^{++}\Lambda A^+_\alpha)
- {1\over 2} D^{++}\mbox{Tr} (\partial^{+\alpha}\Lambda A^+_\alpha) \ ,
  \end{equation}
which is a total derivative with respect to the variables $x^{-}_{\alpha}$ and
$u^\pm_{\alpha'}$.
This allows us to write down the supersymmetric and gauge invariant
action as an integral over the ``1/4 superspace"
$x^{\pm\alpha}, \theta^{+a}, u^\pm_{\alpha'}$:
\begin{equation}\label{31}
S = \int d^4x du d^4\theta^+\; L^{+4}(x,\theta^+,u)\  .   \end{equation}
Obviously,  the variation with respect to the superfields
$A^{++}, A^+_\alpha$ produces the desired field equations
(\ref{0102}),  (\ref{0101}).  Let us now make sure that
the action (\ref{31}) has the correct physical dimension.  Indeed,  the gauge
connections have dimensions $[A^{++}]=0$  (the harmonic variables are
dimensionless),  $[A^+_\alpha]=1$,
so the dimension of the Lagrangian is $[L]=2$.  At the same time, the
superspace measure has dimension $4[dx] + 4[d\theta] = -4 + 2 = -2$,  thus
exactly compensating that of the Lagrangian.  Another property closely
related to the physical dimension is the harmonic weight of the Lagrangian.
By definition, the
harmonic integral in (\ref{31}) would only give a non-vanishing result if the
integrand has zero weight (recall (\ref{6})).  This is indeed true,  since
the weight $+4$ of the Lagrangian is cancelled out by the weight of the
Grassmann measure $d^4\theta^+$.  The last point clearly shows that an action
of this type is only possible in the maximal case $N=4$,  although we could
have written down the Chern-Simons form (\ref{CS}) for any $0\leq N \leq 4$.
The light-cone action of \cite{Parkes,Siegel1,Siegel2} can formally
be written down for $0\leq N<4$ too, although then it requires a dimensionful
coupling constant, which is not natural for a Yang-Mills theory in four
dimensions (see the discussion in \cite{Siegel2,Siegel3}).

Finally, we shall show that the component form of the action
(\ref{31}) is the same as (\ref{33}).  Inserting the Wess-Zumino
gauge (\ref{23}) for $A^{++}$ and the expansion (\ref{25}) of $A^+_{\alpha}$
into
(\ref{31}) and doing the Grassmann integral,  we obtain
\begin{eqnarray}
S &=& \int d^4x du \; \mbox{Tr}\;\left\{ {1\over 2} \phi^{ab} \nabla^{+\alpha}
B^-_{\alpha ab}
+ \chi^{-a} \nabla^{+\alpha} \chi_{\alpha a} + G^{--}{F}^{++}  - C^{-3\alpha}
\partial^{++}{\cal A}^+_\alpha \right. \nonumber \\
&&\left. \ \ \ \ \ - \tau^{--\alpha a} \partial^{++} \chi_{\alpha a}
-{1\over 4} B^{-\alpha ab}\partial^{++}B^-_{\alpha ab} - \phi^{ab}
\chi^\alpha_a
\chi_{\alpha b}\right\} \ ,  \label{32}
 \end{eqnarray}
where $\phi^{ab}=1/2\epsilon^{abcd}\phi_{cd}$, $\nabla^{+\alpha}  =
\partial^{+\alpha}  + [{\cal A}^{+\alpha},  \ ]$, ${F}^{++} =
\partial^{+\alpha}
{\cal A}^+_\alpha + {\cal A}^{+\alpha}{\cal A}^+_\alpha$.
The fields $B^-_{\alpha ab}(x, u)$,  $C^{-3\alpha}(x, u)$,
$\tau^{--\alpha a}(x, u)$ are clearly auxiliary.  They give rise to the
harmonic
equations
$$
\partial^{++}{\cal A}^+_\alpha = 0\ ,  \ \ \partial^{++}
\chi_{\alpha a}=0\ ,  \ \ \partial^{++}B^{-\alpha ab} -
\nabla^{+\alpha} \phi^{ab} =0 \ ,
$$
which allow us to eliminate the harmonic dependence of ${\cal A}^+_\alpha$
and $\chi_{\alpha a}$ and to express $B^{-\alpha ab}$ in terms of
$\phi^{ab}$ (see (\ref{59})).  Afterwards
the harmonic integral in (\ref{32}) becomes trivial and we
arrive at the action (\ref{33}).

\section{Conclusions}

In this paper we presented a harmonic superspace formulation of the
$N$-extended
supersymmetric self-dual Yang-Mills theory in a space with signature $(2,2)$.
We were able to write down an action for the case $N=4$ with
manifest Lorentz invariance and supersymmetry. The most unusual feature is
that the Lagrangian is a Chern-Simons form. In this the $N=4$ SSDYM theory
resembles the $N=3$ SYM theory (signature $(1,3)$) formulated in a harmonic
superspace with harmonics parametrizing the coset ${SU(3)\over U(1)\times
U(1)}$ \cite{N=3}. The main difference is that in the $N=3$ SYM case the
Chern-Simons form is made out of harmonic connections only, whereas in the
$N=4$ SSDYM case we used two space-time and one harmonic one. In both cases
the manifestly supersymmetric formulation greatly facilitates the study of
the quantum properties of the theory.

We remark that a similar formulation exists for the $N=2$ free ``self-dual"
scalar multiplet defined in \cite{Siegel2,Gates}. It can be described by
the {\it anticommuting} harmonic superfields $A_{\alpha i}(x,\theta^+,u)$,
$A^+_{ i}(x,\theta^+,u)$, where $i$ is an index of, e.g.,
an internal symmetry group $SL(2,R)$. The action
is very similar to (\ref{31}):
\begin{equation}\label{s14}
S = \int d^4x du d^2\theta^+ \; \left(A^{+i}\partial^{+\alpha} A_{\alpha i}
-{1\over 2} A^{\alpha i} D^{++} A_{\alpha i}   \right) \ .
  \end{equation}

The most complicated case of a self-dual theory in the space with signature
$(2,2)$ is $N=8$ supergravity. As shown in \cite{Siegel2}, using a light-cone
superspace it can be treated in the same fashion as the self-dual scalar and
Yang-Mills theories. In a future publication we shall present a harmonic
superspace formulation of $N=8$ self-dual supergravity. It will allow us, in
particular, to systematically derive all the supersymmetry transformation laws
of the component fields (they were given in \cite{Siegel2} only partially).

\vskip5mm
{\bf Acknowledgements.} I am very grateful to E. A. Ivanov and
V. I. Ogievetsky for critically reading the manuscript and for a number
of constructive comments
and suggestions. I also thank H. Nishino for stimulating my interest in
the problem. I use this opportunity to express my deepest gratitude to
V. Rittenberg without the friendship and help of whom this and other of my
papers would probably have never been written.

\end{document}